\documentclass{osa-article}
\journal{oe}

\usepackage{color}

\usepackage{graphicx,bm}

\newcommand{\tcr}{\textcolor{black}}

\begin{document}

\title{Efficient construction of witnesses of stellar rank of  nonclassical states of light}

\author{ Jaromír Fiurášek}

\address{Department of Optics, Faculty of Science, Palack\'{y} University, 17. listopadu 1192/12, 77900 Olomouc, Czech Republic}

\begin{abstract}
The stellar hierarchy of quantum states of light classifies the states according to the Fock-state resources  that are required for their generation together with unitary Gaussian operations. States with stellar rank $n$ can be also equivalently referred to as genuinely $n$-photon quantum non-Gaussian states.
Here we present an efficient method for construction of general witnesses of the stellar rank. The number of parameters that need to be optimized in order to determine the witness does not depend on the stellar rank and it scales quadratically with the number of modes. 
We illustrate the procedure by constructing stellar rank witnesses based on pairs of Fock state probabilities and also based on pairs of fidelities with superpositions of coherent states. 
\end{abstract}

\section{Introduction}

Nonclassical states of light such as the squeezed states or Fock states lie at the heart of quantum optics, optical quantum information processing and optical quantum metrology. Given their fundamental importance and a wide range of potential applications, the characterization and classification of nonclassical states of light has attracted 
significant attention. One traditional approach \tcr{to characterization of nonclassical states of optical modes} is based on the $s$-parametrized quasidistributions \cite{Perina1984}, whose special cases are the Wigner function, Husimi function and Glauber-Sudarshan representation. A state is called nonclassical if it is not a coherent state or a statistical mixture of coherent states, i.e. if its Glauber-Sudarshan representation is not an ordinary probability distribution. The degree of nonclassicality can then be quantified by the smallest parameter $s$ for which the $s$-parametrized quasidistribution is not an ordinary probability distribution \cite{Lee1991}. Since the various $s$-parametrized \tcr{quasidistributions} are related via convolution with Gaussian function \cite{Perina1984,Lee1991}, this  approach essentially quantifies how much thermal noise can be added to the state before its nonclassical features vanish. Within this framework, a prominent class of nonclassical states is formed by states with negative Wigner function.

During recent years, a different operationally motivated classification of nonclassical states of light has been proposed, building on the observation that the Gaussian unitary operations form a set of experimentally feasible operations. This lead to the introduction of quantum non-Gaussian states \cite{Filip2011,Walschaers2021,Lachman2022} 
that are defined as states that can not be expressed as pure Gaussian states or their statistical mixture. A number of witnesses of quantum non-Gaussianity have been proposed theoretically \cite{Filip2011,Genoni2013,Lachman2013,Hughes2014,Park2017,Happ2018,Kuhn2018,Fiurasek2021} 
and tested experimentally \cite{Jezek2011,Jezek2012,Straka2014,Baune2014,Straka2018}. Later on, the concept of quantum non-Gaussian 
states has been extended and refined and the so-called genuine  $n$-photon quantum non-Gaussian states have been introduced \cite{Lachman2019}, or, equivalently, the stellar hierarchy of nonclassical states \cite{Chabaud2020}.
A pure single-mode quantum state is said to have stellar rank $n$  if it can be generated by a Gaussian unitary operation from the so-called core state \cite{Menzies2009} that is formed by a finite superposition of Fock states up to $n$, 
\begin{equation}
|\varphi_n\rangle=\hat{U}_G^\dagger \left(\sum_{m=0}^n c_m|m\rangle\right),
\label{statestellarn}
\end{equation}
with nonvanishing amplitude $c_n$.

As shown in Ref. \cite{Chabaud2020}, the Husimi $Q$-function of a pure single-mode state with stellar rank $n$ has  $n$ zeros \tcr{counted with double multiplicity}. A single-mode pure state (\ref{statestellarn}) with stellar rank $n$ can be generated from a Gaussian state by $n$ photon additions or subtractions. 
Specific schemes how to engineer single-mode quantum states of light by repeated photon additions or subtractions  can be found  in Refs. \cite{Dakna1999,Fiurasek2005}. 
A mixed quantum state $\hat{\rho}$ has a stellar rank $n$ if it can be expressed as a mixture of pure states with stellar rank $n$ but not as a mixture of states with lower stellar rank $n-1$ (see Ref. \cite{Chabaud2020} for formal definition).

In order to certify that a quantum state has a stellar rank at least $n$, one can use a suitable witness, in complete analogy with entanglement witnesses \cite{Horodecki2009,Chruscinski2014}. A simplest example is a witness based on fidelity of the certified state $\hat{\rho}$ with some chosen reference state $|w\rangle$ \cite{Chabaud2021}, 
but more general witnesses represented by Hermitian operators $\hat{W}$ can be considered \cite{Lachman2019}. Here we propose and discuss efficient construction of such general witnesses of the stellar rank, thereby generalizing the previous results of Ref. \cite{Chabaud2021} which were obtained for witnesses based on a single state fidelity, i.e. $\hat{W}=| w\rangle\langle w|$. Just like in Ref. ~\cite{Chabaud2021}, in our present approach the number of parameters that need to be optimized does not depend on the stellar rank. 
This should be contrasted to the more computationally demanding approaches based on direct optimization of the coefficients $c_m$ of the core state \cite{Lachman2019,Podhora2021}. 
\tcr{The concept of stellar rank can be extended to multimode states \cite{Chabaud2021,Chabaud2021arXiv} and our procedure for construction of stellar rank witnesses is applicable to both single-mode and multimode systems.}

The rest of the paper is organized as follows. In Section 2 we describe the procedure for efficient construction of general  stellar rank witnesses. In Section 3 we provide two specific examples: witnesses based on Fock state probabilitiies and witnesses based on fidelities with superpositions of coherent states. 
Finally, Section 4 contains a brief summary and conclusions.

\section{Witnesses of stellar rank}

Let us first define the single-mode displacement, squeezing and phase-shift unitary operators
\begin{equation}
\hat{D}(\alpha)=e^{\alpha\hat{a}^\dagger-\alpha^\ast\hat{a}}, \qquad \hat{S}(r)=e^{\frac{r}{2}(\hat{a}^{\dagger 2}-\hat{a}^2)},\qquad \hat{F}(\vartheta)=e^{i\vartheta\hat{n}},
\end{equation}
where $\hat{a}$ and $\hat{a}^\dagger$ denote the annihilation and creation operator, $\hat{n}=\hat{a}^\dagger\hat{a}$ is the photon number operator, $\alpha$ is the (generally complex) coherent displacement, 
$r$ is a real squeezing parameter, and $\vartheta$ denotes the phase shift. Arbitrary single-mode Gaussian unitary operation can be decomposed as a sequence of the above elementary operations \cite{Braunstein2005},
\begin{equation}
\hat{U}_G=\hat{F}(\theta)\hat{D}(\alpha)\hat{S}(r)\hat{F}(\vartheta),
\label{UGdecomposition}
\end{equation}
\tcr{where $\vartheta$  and $\theta$ are independent parameters.}
Note that also other orderings of the operators are possible in Eq. (\ref{UGdecomposition}) and our choice is motivated by technical convenience.

A witness of stellar rank is a Hermitian operator $\hat{W}$ together with the associated thresholds $W_n$. The witness certifies that a quantum state $\hat{\rho}$ has at least stellar rank $n$
if
\begin{equation}
\mathrm{Tr}[\hat{W} \hat{\rho}]> W_n.
\label{witnessinequality}
\end{equation}
Within this hierarchy, the witnesses of quantum non-Gaussian states \cite{Filip2011,Walschaers2021,Lachman2022} i.e. states that cannot be expressed as mixtures of Gaussian states, correspond to $n=1$. 
The threshold $W_n$ can be obtained by maximization over all states $\hat{\rho}_{n-1}$ with stellar rank at most  $n-1$,
\begin{equation}
W_{n}=\tcr{\sup_{\hat{\rho}_{n-1}}} \mathrm{Tr}[\hat{W}\hat{\rho}_{n-1}].
\end{equation}
\tcr{Here and in what follows $\sup$ denotes the supremum.}
\tcr{Since any mixed state $\hat{\rho}_{n-1}$ is a mixture of pure states} $|\varphi_{n-1}\rangle$ with stellar rank at most $n-1$, defined in Eq. (\ref{statestellarn}), \tcr{it suffices  to maximize over these pure states,}
\begin{equation}
W_{n}=\tcr{\sup_{|\varphi_{n-1}\rangle}} \mathrm{Tr}\left[\hat{W}|\varphi_{n-1}\rangle\langle \varphi_{n-1}|\right].
\end{equation}
We now make use of Eq. (\ref{statestellarn}) and denote the core state as $|q_{n-1}\rangle=\sum_{m=0}^{n-1}c_m|m\rangle$. We obtain
\begin{equation}
W_n=\tcr{\sup_{\hat{U}_G, |q_{n-1}\rangle}} \langle q_{n-1}| \hat{U}_G\hat{W} \hat{U}_G^\dagger |q_{n-1}\rangle.
\end{equation}
Let us introduce a projector onto the subspace $\mathcal{H}_n$ of the first $n+1$ Fock states,
\begin{equation}
\hat{\Pi}_{n}=\sum_{m=0}^n |m\rangle\langle m|.
\end{equation}
Since $|q_{n-1}\rangle =\hat{\Pi}_{n-1}|q_{n-1}\rangle$, we have
\begin{equation}
W_n=\tcr{\sup_{\hat{U}_G, |q_{n-1}\rangle}} \langle q_{n-1}| \hat{\Pi}_{n-1}\hat{U}_G\hat{W} \hat{U}_G^\dagger \hat{\Pi}_{n-1} |q_{n-1}\rangle.
\label{Wnbound}
\end{equation}
We now make use of the inequality
\begin{equation}
\langle q_{n-1}| \hat{\Pi}_{n-1}\hat{U}_G\hat{W} \hat{U}_G^\dagger \hat{\Pi}_{n-1} |q_{n-1}\rangle \leq \max \mathrm{eig}\left(\hat{\Pi}_{n-1}\hat{U}_G\hat{W} \hat{U}_G^\dagger \hat{\Pi}_{n-1}\right),
\label{eiginequality}
\end{equation}
that is saturated if $ |q_{n-1}\rangle $ is chosen as the eigenstate corresponding to the largest eigenvalue of the operator
\begin{equation}
\hat{W}_{n,\hat{U}_G}=\hat{\Pi}_{n-1}\hat{U}_G\hat{W} \hat{U}_G^\dagger \hat{\Pi}_{n-1}.
\end{equation}
Note that this is a Hermitian operator  defined on a finite-dimensional subspace hence its eigenvalues can be efficiently calculated numerically, \tcr{or even analytically for $n\leq 4$ by finding the roots of the corresponding characterisitc polynomial.}
When we combine together Eqs.  (\ref{Wnbound}) and (\ref{eiginequality}), we finally obtain
\begin{equation}
W_n=\tcr{\sup_{\hat{U}_G}} \left[ \max \mathrm{eig} \left(\hat{\Pi}_{n-1}\hat{U}_G\hat{W} \hat{U}_G^\dagger \hat{\Pi}_{n-1}\right)\right].
\label{Wnformulafinal}
\end{equation}
Since the phase shift operator $\hat{F}_{\theta}$  maps the subspace $\mathcal{H}_{n-1}$ onto itself,  
\begin{equation}
\hat{F}^\dagger(\theta)\hat{\Pi}_{n-1}\hat{F}(\theta)=\hat{\Pi}_{n-1},
\end{equation}
the operation $\hat{F}(\theta)$ can be omitted in the unitary $\hat{U}_G$ and we can set $\theta=0$ in Eq. (\ref{UGdecomposition})  without any loss of generality. 
The determination of the threshold $W_n$ then involves optimization over four real parameters, namely the squeezing constant $r$, the phase shift $\vartheta$, and the real and imaginary parts of coherent displacement $\alpha$. If the operator $\hat{W}$ is diagonal in Fock basis, 
then also the first phase shift $\hat{F}(\vartheta)$ becomes irrelevant and one needs to optimize over three real parameters only. 

If we choose $\hat{W}=|w\rangle\langle w|$, then we recover the previous results of Chabaud \emph{et al.} \cite{Chabaud2021}. \tcr{ If $\hat{W}=|w\rangle\langle w|$ then} the operator $\hat{W}_{n,\hat{U}_G}$ \tcr{is positive semidefinite} \tcr{and has rank 1,  i.e.}
only one of its eigenvalues can be nonzero. Therefore, in this case
\begin{equation}
\max \mathrm{eig}\left(\hat{W}_{n,\hat{U}_G}\right)=\tcr{\mathrm{Tr}[\hat{W}_{n,\hat{U}_G}]}
\end{equation}
and one needs to maximize $\mathrm{Tr}[\hat{W}_{n,\hat{U}_G}]$ over $\hat{U}_G$. However, the fidelity based witnesses $\hat{W}=|w\rangle\langle w|$ represent only a rather restricted set of all witnesses and more general witnesses exploiting additional data 
can be more powerful. Some specific examples will be presented in the next section. With the present approach, the number of parameters that need to be optimized does not depend on $n$.

Let us now generalize our results to multimode states \tcr{\cite{Chabaud2021,Chabaud2021arXiv}.} We first introduce the $N$-mode Fock states 
\begin{equation}
|\bm{m}\rangle=|m_1\rangle|m_{2}\rangle\cdots|m_k\rangle\cdots|m_N\rangle,
\label{FockNmode}
\end{equation}
where $\bm{m}=(m_1,\ldots,m_{k},\ldots, m_M)$ and
 $|\bm{m}|=\sum_{k=1}^N m_k$ denotes the total number of photons in state (\ref{FockNmode}).
A pure $N$-mode quantum state $|\psi_{n}\rangle$ has a stellar rank $n$ if it can be obtained by an $N$-mode unitary Gaussian operation $\hat{U}_{G,N}$ from a core state formed by superpositions of the multimode Fock states (\ref{FockNmode}) with $|\bm{m}|\leq n$  \cite{Chabaud2021,Chabaud2021arXiv}, 
\begin{equation}
|q_{n,N}\rangle=\sum_{|\bm{m}|\leq n} c_{\bm{m}} |\bm{m}\rangle,
\end{equation}
where at least one of the coefficients $c_{\bm{m}}$ with $|\bm{m}|=n$ is nonzero. We also introduce projector on the subspace of $N$-mode states that do not contain more than $n$ photons in total,
\begin{equation}
\hat{\Pi}_{n,N}=\sum_{|\bm{m}|\leq n} |\bm{m}\rangle\langle \bm{m}|.
\end{equation}

Following the same procedure as above one can show that the thresholds for multimode witness $\hat{W}_N$ can be expressed as
\begin{equation}
W_{N,n}=\tcr{\sup_{\hat{U}_{G,N}} }\left[\max \mathrm{eig} \left(\hat{\Pi}_{n-1,N}\hat{U}_{G,N}\hat{W}_N \hat{U}_{G,N}^\dagger \hat{\Pi}_{n-1,N}\right)\right].
\end{equation}
According to the Bloch-Messiah decomposition \cite{Braunstein2005}, any $N$-mode Gaussian unitary operation $U_{G,N}$ can be decomposed as
\begin{equation}
\hat{U}_{G,N}=\hat{V}_{\mathrm{II}} \left( \bigotimes_{k=1}^N\hat{D}_k(\alpha_k) \hat{S}_k(r_{k})  \right) \hat{V}_{\mathrm{I}}.
\end{equation}
Here $\hat{V}_{\mathrm{I}}$ and $\hat{V}_{\mathrm{II}}$ are unitary operators that describe the action of  general $N$-mode passive linear optical interferometers, and $N$ single-mode squeezers and displacers are sandwiched in between those two interferometers. 
Since $\hat{V}_{\mathrm{II}}$ commutes with $\hat{\Pi}_{n,N}$, it can be omitted in the optimization, just like the phase-shift operator $\hat{F}(\theta)$ in the single-mode case.

\tcr{To provide additional interpretation for the stellar rank witnesses, we can consider the trace distance \cite{Chabaud2021Wigner} between the state $\hat{\rho}$ and the set  $\mathcal{S}_{n-1}$ of states $\hat{\sigma}$ that have stellar rank at most $n-1$,
\begin{equation}
D_{n}(\hat{\rho})=\inf_{\hat{\sigma}\in \mathcal{S}_{n-1}} \mathrm{Tr}\left[|\hat{\rho}-\hat{\sigma}|\right].
\end{equation}
Specifically, as shown in Ref. \cite{Chabaud2021Wigner} in the context of witnesses of Wigner function negativity, the value of the witness can provide a lower bound on $D_{n}(\hat{\rho})$. In particular, if  $\hat{W}$ is a witness operator that satisfies 
\begin{equation}
0\leq \hat{W}\leq \hat{I},
\label{WDconditions}
\end{equation}
 then
\begin{equation}
D_{n}(\hat{\rho}) \geq \mathrm{Tr}\left[\hat{W}\hat{\rho}\right]-W_{n}.
\label{Dbound}
\end{equation}
 The proof of the inequality (\ref{Dbound}) parallels  the proof of Lemma 1 for witnesses of Wigner function negativity presented in Ref. \cite{Chabaud2021Wigner}. Namely, the lower bound (\ref{Dbound}) follows from  the following set of inequalities that hold for any $\hat{\sigma}\in \mathcal{S}_{n-1}$,
\begin{equation}
\mathrm{Tr}[|\hat{\rho}-\hat{\sigma}|]\geq \mathrm{Tr}[\hat{W}|\hat{\rho}-\hat{\sigma}|]\geq \mathrm{Tr}[\hat{W}(\hat{\rho}-\hat{\sigma})] \geq \mathrm{Tr}[\hat{W}\hat{\rho}]-W_{n}.
\end{equation}
Here the last inequality holds because $\mathrm{Tr}[\hat{W}\hat{\sigma}]\leq W_n$ for any $\hat{\sigma}\in \mathcal{S}_{n-1}$. If the operator $\hat{W}$ is bounded, then it can be converted by a linear transformation into a new equivalent witness operator that satisfies the conditions (\ref{WDconditions}):
$\hat{W}\rightarrow a\hat{W}+b\hat{I}$, where $a$  and $b$ are suitable nonnegative constants. The witness thresholds transform accordingly, $W_n\rightarrow aW_n+b$ .}

\section{Examples of stellar-rank witnesses}

\subsection{Witnesses based on Fock states}

\tcr{Let} us first consider witnesses based on Fock state probabilities $p_m=\langle m|\hat{\rho}|m\rangle$. The witness operator becomes diagonal in Fock basis,
\begin{equation}
\hat{W}=\sum_{m=0}^\infty w_m |m\rangle\langle m|,
\label{WFockgeneral}
\end{equation}
and is specified by real coefficients $w_m$. \tcr{Quantum non-Gaussianity criteria that can be represented by Fock-state diagonal  witnesses (\ref{WFockgeneral}) have been investigated in a number of works \cite{Filip2011,Jezek2011,Lachman2013,Straka2014,Straka2018,Lachman2019,Fiurasek2021}, 
and the witnesses (\ref{WFockgeneral}) can also be used to certify Wigner function negativity \cite{Chabaud2021Wigner}. }
As already pointed out above, since (\ref{WFockgeneral}) is invariant with respect to phase shifts $\hat{F}(\vartheta)$, determination of the witness thresholds $W_n$ requires optimization over three real parameters.
To provide a specific example, we construct witnesses based on pairs of Fock-state probabilities $p_j$ and $p_k$. We can define the corresponding operators $\hat{W}$ as
\begin{equation}
\hat{W}_{jk}=\cos\omega |j\rangle\langle j |+\sin\omega  |k\rangle\langle k |.
\end{equation}
For each value of $\omega$ and chosen stellar rank $n$ we numerically determine the optimal Gaussian unitary operation $\hat{U}_G(\omega,n)$ and the optimal core state $|q_{n-1}(\omega)\rangle$. 
The matrix elements of a Gaussian unitary operator in Fock basis required for the calculations can be analytically expressed as a linear combination of products of Hermite polynomials \cite{Kral1999},
\begin{eqnarray}
\langle k|\hat{U}_G |m \rangle
& =&e^{-i k\theta} e^{i m \vartheta} \left(\frac{k! m!}{\mu}\right)^{1/2}\left(\frac{\nu}{2\mu}\right)^{(k+m)/2} \exp\left(-\frac{|\alpha|^2}{2}+\frac{\nu}{2\mu}\alpha^{\ast 2}\right) \nonumber \\
& & \times \sum_{j=0}^{\min(k,m)}\frac{2^j i^{k-j} (\nu)^{-j}}{j!(m-j)!(k-j)!} H_{m-j}\left(\frac{-\alpha^\ast}{\sqrt{2\mu\nu}}\right) H_{k-j}\left(\frac{\mu\alpha-\nu\alpha^\ast}{\sqrt{-2\mu\nu}}\right), \nonumber \\
\label{Fockmatrixelements}
\end{eqnarray}
where $\mu=\cosh r$ and $\nu=\sinh r$.

\begin{figure}[!t]
	\centering
	\centerline{\includegraphics[width=0.99\linewidth]{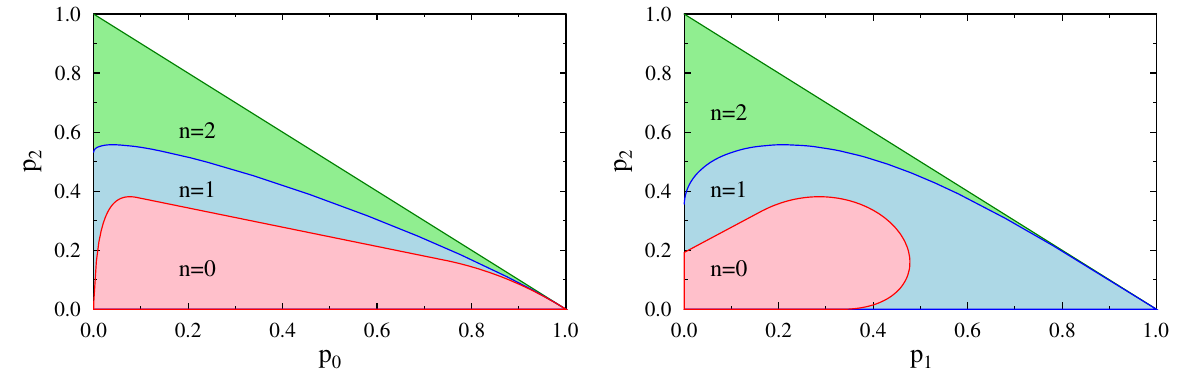}}
\caption{
		Visualization of stellar rank witnesses based on pairs of Fock state probabilities $p_0$ and $p_2$ (a), and $p_1$ and $p_2$ (b). The minimal stellar rank $n$ that is certified 
by a given pair of probabilities is specified as a label of each coloured area. Witnesses can be constructed as tangent lines to the chosen boundary curve between two differently coloured areas. }
	\label{fig:Fock}
\end{figure} 

To visualize the obtained witnesses, we calculate the pair of probabilities 
\begin{equation}
\tcr{p_j(\omega,n)=|\langle q_{n-1}(\omega)|\hat{U}_G(\omega,n)|j\rangle|^2, \qquad  p_k(\omega,n)=|\langle q_{n-1}(\omega)|\hat{U}_G(\omega,n)|k\rangle|^2.}
\label{pjkparametric}
\end{equation}
These probability pairs form extremal points of a convex set in a plane, and one should construct their convex hull, e.g. by applying the gift wrapping algorithm \cite{Jarvis1973}. It turns out that for the examples considered below the boundary of the convex hull is either given by a continuous curve parametrically specified by Eq. (\ref{pjkparametric}), or, in case of discontinuity, one simply  connects the neighboring extremal points with a straight line.  
Concrete examples are provided in Fig.~1 for the probability pairs $(p_0,p_2)$ and  $(p_1, p_2)$. A witness can then be represented by a tangent line to the boundary curve at a chosen point on that curve. 
The witnesses certify that the quantum state with Fock state probabilities $p_j$ and $p_k$ has a stellar rank at least $n$ if the point $(p_j,p_k)$ lies in the corresponding area in the figure. 
We emphasize that the primary outputs of the optimization are the witness thresholds and for any chosen witness the stellar rank $n$ can be directly certified by checking the inequality (\ref{witnessinequality}) even without resorting to the constructions plotted in Fig.~1. 

If a witness is constructed from  $N$ Fock state probabilities, then the witness defines a hyperplane in an $N$-dimensional space that splits this space into two halves. Such multidimensional stellar rank witnesses would be analogous 
to the criteria of quantum state nonclassicality derived in Ref. \cite{Innocenti2022}.
One can also deal with witnesses that involve an infinite number of Fock state probabilities. For instance, qantum non-Gaussianity criteria based on $p_0$  and a vacuum probability $q_0(T)$ 
after transmission through a lossy channel with transmittance $T$ depend on all $p_m$, since $q_0(T)=\sum_{m=0}^\infty (1-T)^m p_m$ \cite{Lachman2013,Fiurasek2021}.
 \tcr{Note that $q_0(T)$ can be also expressed in terms of the overlap of the state $\hat{\rho}$ with a thermal state $\hat{\tau}_{\mathrm{th}}(\bar{n})$ with mean photon number $\bar{n}=(1-T)/T$ \cite{Hlousek2021},
\begin{equation}
q_0(T)=\mathrm{Tr} \left[(1-T)^{\hat{n}} \hat{\rho}\right]=(1+\bar{n}) \mathrm{Tr}[\hat{\tau}_{\mathrm{th}}(\bar{n}) \hat{\rho}].
\label{q0T}
\end{equation}
The photon number distribution $p_m$ can be directly measured with transition-edge sensors \cite{Lita2008,Morais2020,Harder2016,Eaton2022} or it can be inferred from measurements with multiplexed detectors, where the incoming signal is spatially or temporally 
split among several binary on-off detectors that distinguish the presence and absence of photons \cite{Paul1996,Sperling2012,Achilles2003,Fitch2003,Rehacek2003,Divochiy2008,Zhu2018,Hlousek2019,Tiedau2019,Cheng2022}. }

\tcr{
For the multiplexed partially photon-number resolving detectors, it is more natural to formulate the witnesses directly in terms of the probabilities of the various observed $k$-fold coincidence events \cite{Straka2018,Lachman2019}, i.e. in terms of the click statistics $c_k$, which is given by
 linear combinations of the photon number probabilities, $c_k=\sum_{m=0}^\infty C_{km} p_m$. Here $C_{km}\geq 0$ specifies the response of the multiplexed detector to input Fock state $|m\rangle$ and the index $k$ labels the various coincidence events that can be observed. 
While for a fully balanced detector with $M$ channels there will be  $M+1$ different coincidence events due to degeneracy, this number will increase for an unbalanced detector. For example, a  simple two-channel detector formed by an unbalanced beam splitter with transmittance $T$ with
 an on-off detector placed at each output port will provide $4$ different click events:  none of the detectors clicks, both detectors click, only the first detector clicks, and only the second detector clicks \cite{Rehacek2003}. 
Each binary on-off detector can be parametrized by its detection efficiency $\eta$ and  dark count rate $R_D$,
and it can be modeled by a lossy channel $\mathcal{L}$ with transmittance $\eta$ followed by detector described by a binary POVM $\hat{\Pi}_0=(1-R_D)|0\rangle\langle 0|$ and $\hat{\Pi}_1=\hat{I}-(1-R_D)|0\rangle\langle 0|$.  
Interestingly, the probability $c_k$ of certain coincidence event in a multiplexed detector composed of such elementary detectors  can be expressed as a linear combination of vacuum probabilities $q_0(T)$ with several different $T$ \cite{Lachman2022}. For instance, for a balanced $M$-channel  multiplexed 
detector with $M$ identical on-off detectors the probability of simultaneous click of a specific subset of $m$ detectors can be expressed as \cite{Hlousek2019,Lachman2022}
\begin{equation}
c_m= \sum_{j=0}^m  { m \choose j} (-1)^j (1-R_D)^{M-m+j}q_0\left( \eta \frac{M-m+j}{M}\right).
\end{equation}
Taking into account Eq. (\ref{q0T}) we observe that for witnesses formed by linear combinations of the probabilities of coincidence counts $c_k$ the operator $\hat{W}$ can be expressed as a linear combination of density matrices of thermal states. This observation can facilitate the calculation of the witness thresholds. 
Any Gaussian unitary $\hat{U}_G$ transforms  thermal states onto mixed Gaussian states whose density matrix elements can be evaluated analytically.}

\tcr{
The single-photon detectors respond to the total number of photons that impinge on the detector within the measurement time. In the experiments, where one cannot guarantee the single-mode nature of the detected signal, and where e.g. several temporal or frequency modes can be present, the single-mode witnesses become 
insufficient. In such case one should consider multimode witnesses \cite{Fiurasek2021} and the certification of stellar rank can become more challenging and involved. }

\subsection{Witnesses based on Schr\"{o}dinger cat-like states}

As a second example we consider witnesses based on coherent superpositions of two coherent states,
\begin{equation}
|\beta_{+}\rangle=\frac{1}{\sqrt{2 \mathcal{N}_{+}}}(|\beta\rangle+|-\beta\rangle), \qquad |\beta_{-}\rangle=\frac{1}{\sqrt{2\mathcal{N}_{-} }}(|\beta\rangle-|-\beta\rangle),
\end{equation}
where $\mathcal{N}_{\pm}(\beta)=1\pm\exp(-2|\beta|^2)$. In particular, we shall consider witnesses 
\begin{equation}
\hat{W}=\cos(\omega)|\beta_{-}\rangle\langle \beta_{-}|+ \sin(\omega) |\beta_{+}\rangle\langle \beta_{+}|,
\label{Wcat}
\end{equation}
where $\omega$ is a real parameter. In experiments, state preparation and transmission is usually affected by losses. A lossy channel with amplitude transmittance $t$ 
transfers the pure states $|\beta_{\pm}\rangle$ onto mixed states, which can be expressed as statistical mixtures of $|t\beta_{+}\rangle$ and $|t\beta_{-}\rangle$. For instance, state $|\beta_{+}\rangle$ is transferred to 
\begin{equation}
\hat{\rho}_{+}= \frac{1}{2\mathcal{N}_{+}(\beta)} \left[|t\beta\rangle\langle t\beta|+|-t\beta\rangle\langle -t\beta|+e^{-2(1-t^2)|\beta|^2}\left(|-t\beta\rangle\langle t\beta|+|t\beta\rangle\langle -t\beta|\right)\right].
\end{equation}
This can be equivalently written as
\begin{equation}
\hat{\rho}_{+}=p_{+}|t\beta_{+}\rangle \langle t\beta_{+}|+(1-p_{+}) |t\beta_{-}\rangle \langle t\beta_{-}|,
\label{rhocatlossy}
\end{equation}
where
\begin{equation}
p_{+}=\frac{1}{2}+\frac{1}{2} \frac{e^{-2t^2|\beta|^2}+e^{-2(1-t^2)|\beta|^2}}{1+e^{-2|\beta|^2}}.
\end{equation}
The structure of the density operator (\ref{rhocatlossy}) motivates the construction of the witness (\ref{Wcat}).

\begin{figure}[!t]
	\centering
	\centerline{\includegraphics[width=0.6\linewidth]{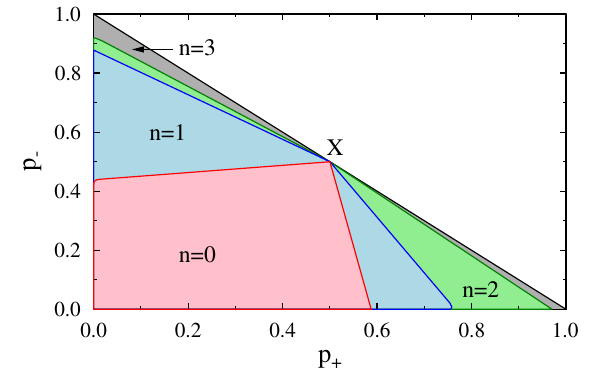}}
	\caption{
		Certification of stellar rank based on fidelities $p_+$ and $p_{-}$ with even and odd coherent states $|\beta_{+}\rangle$ and $|\beta_{-}\rangle$, respectively. Here $\beta=2$ is chosen and a stellar rank at least $n$ is certified if the probability pair $(p_{+},p_{-})$ lies in the corresponding area. Note that the 
even and odd coherent states have infinite stellar rank \cite{Chabaud2020}. In the figure we plot only the boundary curves up to $n=3$, as for higher stellar ranks the areas become progressively smaller and smaller. } 
	\label{fig:cat}
\end{figure} 

In order to calculate the matrix elements of the transformed operator $\hat{\Pi}_{n-1}\hat{U}_G\hat{W}\hat{U}_G^\dagger \hat{\Pi}_{n-1}$ in Fock basis,
we make use of the following identity
\begin{equation}
\hat{D}(\alpha)\hat{S}(r)\hat{F}(\vartheta)|\beta\rangle=e^{(\alpha\tilde{\beta}^\ast-\alpha^\ast\tilde{\beta})/2} \hat{D}(\alpha+\tilde{\beta})\hat{S}(r)|0\rangle,
\end{equation}
where $\tilde{\beta}=e^{i\vartheta} \beta \cosh r  +e^{-i\vartheta}\beta^\ast \sinh r$, together with the expression (\ref{Fockmatrixelements}) for matrix elements of unitary Gaussian operations. 

Results of numerical optimization of the witnesses (\ref{Wcat}) 
are shown in Fig.~2 for $\beta=2$. Similarly as in the previous subsection, we display areas of probability pairs $p_{+}=\langle \beta_+|\hat{\rho}|\beta_{+}\rangle$ and $p_{-}=\langle \beta_-|\hat{\rho}|\beta_{-}\rangle$ that certify certain stellar rank $n$. 
 For the considered example with $\beta=2$, the constructed convex hulls are essentially convex quadrilaterals. 
For the considered $\beta$ 
the two coherent states $|\beta\rangle$ and $|-\beta\rangle$ become practically orthogonal and a coherent state $|\beta\rangle$ has fidelity very close to $1/2$ with both $|\beta_{+}\rangle$ and $\beta_{-}\rangle$, which corresponds to the point $X$ in Fig.~2.
By contrast, in the limit $\beta\rightarrow 0$ we get $|\beta_{+}\rangle=|0\rangle$ and $|\beta_{-}\rangle=|1\rangle$ and we recover the previously derived non-Gaussianity witnesses based on probabilities of vacuum and single-photon states \cite{Filip2011,Jezek2011}. 
Consequently, for small $\beta$ the witnesses (\ref{Wcat})  can reasonably certify only stellar rank $n=1$. 

In experiments, the fidelities $p_{\pm}$  with the odd and even coherent states can be directly determined by averaging suitable sampling functions over quadrature data measured with balanced homodyne detector \cite{Fiurasek2013}. Alternatively, the fidelities can be obtained from tomographically reconstructed density matrix.
\tcr{If we posses full characterization of the state, obtained e.g. through the quantum state tomography,  then we can take the eigenstates of the density matrix $\hat{\rho}$ that correspond to the largest eigenvalues of $\hat{\rho}$ and build witnesses based on linear combinations of projectors onto these dominant eigenstates, 
thus effectively probing the state properties in the relevant subspace. }

\section{Conclusions \tcr{and outlook}}
In summary, we have presented an efficient method for numerical construction of general witnesses of stellar rank of nonclassical states of light. These witnesses enable detailed characterization and classification of nonclassical quantum states. 
The stellar rank has a clear operational meaning \tcr{because it quantifies the Fock state resources that  are required, together with Gaussian operations, to prepare a given state.}
For single-mode states the stellar rank \tcr{equivalently} specifies the minimum number of photon additions or subtractions required to generate a given state from input Gaussian state, \tcr{and  for multimode states it gives a lower bound on the number of required photon additions or subtractions \cite{Chabaud2021arXiv}.}
Determination of single-mode stellar rank witnesses requires optimization over four real parameters irrespective of the targeted stellar rank, which is feasible with current computational resources. 
The calculations become more challenging for multimode witnesses, where the number of parameters scales quadratically with the number of modes. For low stellar ranks, where analytical expressions for eigenvalues of operator $\hat{W}_{n,\hat{U}_G}$ exist, one can attempt to combine numerical and analytical approaches based on formulation 
of analytical extremal equations for the optimized parameters \cite{Filip2011}. However, for higher \tcr{stellar} ranks fully numerical approaches seem unavoidable. Each constructed witness $\hat{W}$ in fact provides  a whole class of witnesses that are generated from $\hat{W}$ by \tcr{arbitrary 
unitary Gaussian operation $\hat{V}_G$, because} the witness thresholds $W_n$ are the same for all \tcr{$\hat{V}_G \hat{W}\hat{V}_G^\dagger$} and do not depend on \tcr{$\hat{V}_G$}. \tcr{This can be easily seen from Eq. (\ref{Wnformulafinal}) where for any chosen $\hat{V}_G$ this additional Gussian unitary can be simply 
absorbed into the varying $\hat{U}_G$, $\hat{U}_G \hat{V}_{G} \rightarrow \hat{U}_G$.}
The witnesses \tcr{of stellar rank} are applicable to any physical system where nonclassical states of harmonic oscillators may emerge. Besides optics, this includes for instance quantum optomechanics or motional states of trapped ions \cite{Podhora2021}.

\tcr{The Majorana stellar representation was originally introduced for spin states \cite{Majorana1932,Bengtsson2006,Bruno2012,Goldberg2020}. A pure quantum state of spin $J$ can be represented by $2J$ points on the Bloch sphere, called the stars. 
The position of stars is determined by the roots of a certain polynomial, and corresponds to zeros of the Husimi function of the spin state \cite{Bengtsson2006,Goldberg2020}. 
A state of spin $J$ can be considered nonclassical if it cannot be expressed as a mixture of spin coherent states \cite{Giraud2008}. The spin coherent states are the only pure states for which all the $2J$ stars coincide, while for states that can be considered as highly nonclassical the stars are spread around the Bloch sphere \cite{Bouchard2017,Goldberg2020}. The Hilbert space of the spin $J$ system is spanned by  $2J+1$ basis states $|J,m\rangle$, that are the eigenstates of $\hat{J}^2$ and $\hat{J}_z$. 
A possible definition of stellar rank for spin system could be based on considering the core states of the form $\sum_{k=0}^n c_k|J,J-k\rangle$ with $|J,J-k\rangle$ replacing the vacuum and Fock states in Eq. (\ref{statestellarn}), 
and the Gaussian unitaries would be replaced by $SU(2)$ transformations, i.e. by rotations of the Bloch sphere in the stellar representation. Such pure spin states of stellar rank $n$ would have Majorana representation where all the stars except $n$ coincide. Witnesses of such spin stellar rank could be constructed by the procedure described in Section 2, with the optimization over $\hat{U}_G$ replaced by optimization over the $SU(2)$ transformations (Bloch sphere rotations). However, a detailed analysis of this is beyond the scope of the present paper and it remains to be assessed how physically relevant and useful this notion could be for the spin systems. }

\section*{Funding}
 Czech Science Foundation (21-23120S).

\section*{Disclosure}
The author declares that there are no conflicts of interest related to this article.

\section*{Data availability}
Data underlying the results presented in this study are not publicly available at this time but may be obtained from the author upon a reasonable request.


\begin{thebibliography}{99}

\bibitem{Perina1984}
J. Pe\v{r}ina, ``Quantum Statistics of Linear and Nonlinear Optical Phenomena'' (Dordrecht: Reidel, 1984).

\bibitem{Lee1991}
C.T. Lee, ``Measure of the nonclassicality of nonclassical states,'' Phys. Rev. A \textbf{44}(5), R2775-R2778 (1991).


\bibitem{Filip2011}
R. Filip and L. Mi\v{s}ta, Jr., ``Detecting Quantum States with a Positive Wigner Function beyond Mixtures of Gaussian States,'' Phys. Rev. Lett. \textbf{106}(20), 200401 (2011).


\bibitem{Walschaers2021}
M. Walschaers, ``Non-Gaussian Quantum States and Where to Find Them,'' PRX Quantum \textbf{2}(3), 030204 (2021).

\bibitem{Lachman2022}
L. Lachman and R. Filip, ``Quantum non-Gaussianity of light and atoms,'' Progress in Quantum Electronics \textbf{83}, 100395 (2022).


\bibitem{Genoni2013}
M.G. Genoni, M.L. Palma, T. Tufarelli, S. Olivares, M. S. Kim, and M.G.A. Paris, ``Detecting quantum non-Gaussianity via the Wigner function,'' Phys. Rev. A \textbf{87}(6), 062104 (2013).


\bibitem{Lachman2013}
L. Lachman and R. Filip, ``Robustness of quantum nonclassicality and non-Gaussianity of single-photon states in attenuating channels,'' Phys. Rev. A \textbf{88}(6), 063841 (2013).


\bibitem{Hughes2014}
C. Hughes, M.G. Genoni, T. Tufarelli, M.G.A. Paris, and M.S. Kim, ``Quantum non-Gaussianity witnesses in phase space,'' Phys. Rev. A \textbf{90}(1), 013810 (2014).


 \bibitem{Park2017}
J. Park, Y. Lu, J. Lee, Y. Shen, K. Zhang, S. Zhang, M.S. Zubairy, K. Kim, and H. Nha, ``Revealing nonclassicality beyond Gaussian states via a single marginal distribution,'' PNAS \textbf{114}(5), 891-896 (2017). 

\bibitem{Happ2018}
L. Happ, M.A. Efremov, H. Nha, and W.P. Schleich, ``Sufficient condition for a quantum state to be genuinely quantum non-Gaussian,'' New J. Phys. \textbf{20}(2), 023046 (2018).


\bibitem{Kuhn2018}
B. K\"{u}hn and W. Vogel, ``Quantum non-Gaussianity and quantification of nonclassicality,'' Phys. Rev. A \textbf{97}(5), 053823 (2018).


\bibitem{Fiurasek2021}
J. Fiurášek, L. Lachman and R. Filip, ``Quantum non-Gaussianity criteria based on vacuum probabilities of original and attenuated state,'' New J. Phys. \textbf{23}(7), 073005 (2021).


\bibitem{Jezek2011}
M. Je\v{z}ek, I. Straka, M. Mi\v{c}uda, M. Du\v{s}ek, J. Fiur\'{a}\v{s}ek, and R. Filip, ``Experimental Test of the Quantum Non-Gaussian Character of a Heralded Single-Photon State,''  Phys. Rev. Lett. \textbf{107}(21), 213602 (2011).

\bibitem{Straka2014} 
I. Straka, A. Predojevi\.{c}, T. Huber, L. Lachman, L. Butschek, M. Mikov\'{a}, M. Mi\v{c}uda, G.S. Solomon, G. Weihs, M. Je\v{z}ek, and R. Filip, ``Quantum non-Gaussian Depth of Single-Photon States,''  Phys. Rev. Lett. \textbf{113}(22), 223603 (2014).

\bibitem{Jezek2012}
M. Ježek, A. Tipsmark, R. Dong, J. Fiurášek, L. Mišta, Jr., R. Filip, and U. L. Andersen. ``Experimental test of the strongly nonclassical character of a noisy squeezed single-photon state,'' Phys. Rev. A \textbf{86}(4), 043813 (2012).

\bibitem{Baune2014}
C. Baune, A.l Sch\"{o}nbeck, A. Samblowski, J. Fiur\'{a}\v{s}ek, and R. Schnabel,  ``Quantum non-Gaussianity of frequency up-converted single photons,''  Opt. Express \textbf{22}(19),  22808-22816 (2014). 

\bibitem{Straka2018}
I. Straka, L. Lachman, J. Hlou\v{s}ek, M. Mikov\'{a}, Michal Mi\v{c}uda, M. Je\v{z}ek, and R. Filip, Quantum non-Gaussian multiphoton light,''  npj Quant. Inf. \textbf{4}, 4 (2018).


\bibitem{Lachman2019}
L. Lachman, I. Straka, J. Hlou\v{s}ek, M. Je\v{z}ek, and R. Filip, ``Faithful Hierarchy of Genuine n-Photon Quantum Non-Gaussian Light,'' Phys. Rev. Lett. \textbf{123}(4), 043601 (2019).


\bibitem{Chabaud2020}
U. Chabaud, D. Markham, and F. Grosshans, ``Stellar Representation of Non-Gaussian Quantum States,'' Phys. Rev. Lett. \textbf{124}(6), 063605 (2020).


\bibitem{Menzies2009}
D. Menzies and R. Filip, ``Gaussian-optimized preparation of non-Gaussian pure states,'' Phys. Rev. A \textbf{79}(1), 012313 (2009).


\bibitem{Dakna1999}
M. Dakna, J. Clausen, L. Kn\"{o}ll, and D.-G. Welsch, ``Generation of arbitrary quantum states of traveling fields,'' Phys. Rev. A \textbf{59}(2), 1658-1661 (1999).

\bibitem{Fiurasek2005}
J. Fiurášek, R. García-Patrón, and Nicolas J. Cerf, ``Conditional generation of arbitrary single-mode quantum states of light by repeated photon subtractions,'' Phys. Rev. A \textbf{72}(3), 033822 (2005).


\bibitem{Horodecki2009}
R. Horodecki, P. Horodecki, M. Horodecki and K. Horodecki, ``Quantum entanglement,'' Rev. Mod. Phys. \textbf{81}(2), 865-942 (2009).

\bibitem{Chruscinski2014}
D. Chru\'{s}ci\'{n}ski and G. Sarbicki, ``Entanglement witnesses: construction, analysis and classification,'' J. Phys. A: Math. Theor. \textbf{47}(48), 483001 (2014).


\bibitem{Chabaud2021}
U. Chabaud, G. Roeland, M. Walschaers, F.c Grosshans, V. Parigi, D. Markham, and N. Treps, ``Certification of Non-Gaussian States with Operational Measurements,'' PRX Quantum \textbf{2}(2), 020333 (2021).

\bibitem{Podhora2021}
L. Podhora, L. Lachman, T. Pham, A. Le\v{s}und\'{a}k, O. \v{C}\'{\i}p, L. Slodi\v{c}ka, and R. Filip, \tcr{``Quantum Non-Gaussianity of Multiphonon States of a Single Atom,'' Phys. Rev. Lett. \textbf{129}(1), 013602  (2022).}



\bibitem{Chabaud2021arXiv}
\tcr{U. Chabaud and S. Mehraban, ``Holomorphic Quantum Computing,'' arXiv:2111.00117.}


\bibitem{Braunstein2005}
S.L. Braunstein, ``Squeezing as an irreducible resource,'' Phys. Rev. A \textbf{71}(5), 055801 (2005).


\bibitem{Chabaud2021Wigner}
\tcr{U. Chabaud, P.-E. Emeriau, and F. Grosshans, ``Witnessing Wigner Negativity,'' Quantum \textbf{5}, 471 (2021).}


\bibitem{Kral1999}
P. Kr\'{a}l, ``Displaced and Squeezed Fock States,'' J. Mod. Opt. \textbf{37}(5), 889-917 (1990).

\bibitem{Jarvis1973}
R. A. Jarvis, ``On the identification of the convex hull of a finite set of points in the plane,'' Inf. Process. Lett. \textbf{2}(1), 18–21  (1973). 


\bibitem{Innocenti2022}
L. Innocenti, L. Lachman, and R. Filip, ``Nonclassicality detection from few Fock-state probabilities,''  npj Quantum Information \textbf{8}, 30 (2022). 


\bibitem{Hlousek2021}
\tcr{J. Hlou\v{s}ek, M. Je\v{z}ek, and J. Fiur\'{a}\v{s}ek, Direct Experimental Certification of Quantum Non-Gaussian Character and Wigner Function Negativity of Single-Photon Detectors, Phys. Rev. Lett. \textbf{126}(4), 043601 (2021).}




\bibitem{Eaton2022}
\tcr{M. Eaton, A. Hossameldin, R.J. Birrittella, P.M. Alsing, C.C. Gerry, C. Cuevas, H. Dong, and O. Pfister, ``Resolving 100 photons and quantum generation of unbiased random numbers,' arXiv:2205.01221.}


\bibitem{Lita2008}
\tcr{A. E. Lita, A. J. Miller, and S. W. Nam, ``Counting near-infrared single-photons with 95\% efficiency,''  Opt. Expr. \textbf{16}(5), 3032--3040 (2008).}

\bibitem{Morais2020}
L. A. Morais, T.Weinhold, M. P. de Almeida, A. Lita, T. Gerrits, S. W. Nam, A. G. White, and G. Gillett, ``Precisely determining photon-number in real-time,'' arXiv:2012.10158 (2020).

\bibitem{Harder2016}
\tcr{G. Harder, T.J. Bartley, A.E. Lita, S.W. Nam, T. Gerrits, and C. Silberhorn, ``Single-Mode Parametric-Down-Conversion States with 50 Photons as a Source for Mesoscopic Quantum Optics,'' Phys. Rev. Lett. \textbf{116}(14), 143601 (2016).}

\bibitem{Paul1996}
\tcr{H. Paul, P. Törmä, T. Kiss, and I. Jex, ``Photon Chopping: New Way to Measure the Quantum State of Light,''  Phys. Rev. Lett. \textbf{76}(14), 2464-2467 (1996).}

\bibitem{Sperling2012}
\tcr{J. Sperling, W. Vogel, and G. S. Agarwal, ``True photocounting statistics of multiple on-off detectors,'' Phys. Rev. A \textbf{85}(2), 023820 (2012).}

\bibitem{Achilles2003}
\tcr{D. Achilles, C. Silberhorn, C. \'{S}liwa, K. Banaszek, and I. A. Walmsley, ``Fiber-assisted detection with photon number resolution,'' Opt. Lett. \textbf{28}(23), 2387--2389 (2003).}

\bibitem{Fitch2003}
\tcr{M. J. Fitch, B. C. Jacobs, T. B. Pittman, and J. D. Franson, ``Photon-number resolution using time-multiplexed single-photon detectors,'' Phys. Rev. A \textbf{68}(4), 043814 (2003).}

\bibitem{Rehacek2003}
\tcr{ J. \v{R}eh\'{a}\v{c}ek, Z. Hradil, O. Haderka, J. J. Pe\v{r}ina, and M. Hamar, ``Multiple-photon resolving fiber-loop detector,'' Phys. Rev. A \textbf{67}(6), 061801(R) (2003).}

\bibitem{Tiedau2019}
\tcr{J. Tiedau, E. Meyer-Scott, T. Nitsche, S. Barkhofen, T. J. Bartley, and C. Silberhorn, ``A high dynamic range optical detector for measuring single photons and bright light,'' Opt. Express \textbf{27}(1), 1-15 (2019).}

\bibitem{Divochiy2008}
\tcr{A. Divochiy, F. Marsili, D. Bitauld, A. Gaggero, R. Leoni, F. Mattioli, A. Korneev, V. Seleznev, N. Kaurova, O. Minaeva, G. Gol’tsman, K. G. Lagoudakis, M. Benkhaoul, F. L\'{e}vy, and A. Fiore, 
``Superconducting nanowire photon-number-resolving detector at telecommunication wavelengths,'' Nature Photon. \textbf{2}(5), 302-306 (2008).}

\bibitem{Zhu2018}
\tcr{D. Zhu, Q.-Y. Zhao, H. Choi, T.-J. Lu, A. E. Dane, D. Englund, and K. K. Berggren, A scalable multi-photon coincidence detector based on superconducting nanowires, Nat. Nanotechnol. \textbf{13}(7), 596-601 (2018).}

\bibitem{Hlousek2019}
\tcr{J.~Hlou{\v{s}}ek, M.~Dudka, I.~Straka, and M.~Je{\v{z}}ek, ``Accurate  detection of arbitrary photon statistics,'' Phys. Rev.  Lett. \textbf{123}(15), 153604  (2019).}

\bibitem{Cheng2022}
\tcr{R. Cheng, Y. Zhou, S. Wang, M. Shen, T. Taher, H.X. Tang, ``Unveiling photon statistics with a 100-pixel photon-number-resolving detector'', arXiv:2206.13753. }


\bibitem{Fiurasek2013}
J. Fiur\'{a}\v{s}ek and M. Je\v{z}ek, ``Witnessing negativity of Wigner function by estimating fidelities of catlike states from homodyne measurements,'' Phys. Rev. A \textbf{87}(6), 062115 (2013).

\bibitem{Majorana1932}
\tcr{E. Majorana, “Atomi orientati in campo magnetico variabile,” Nuovo Cimento 9, 43–50 (1932).}

\bibitem{Bengtsson2006}
\tcr{I. Bengtsson and K. \.{Z}yczkowski, \emph{Geometry of Quantum States: An Introduction to Quantum Entanglement}  (Cambridge University Press, 2006).}

\bibitem{Bruno2012}
\tcr{P. Bruno, ``Quantum Geometric Phase in Majorana’s Stellar Representation: Mapping onto a Many-Body Aharonov-Bohm Phase,'' Phys. Rev. Lett. \textbf{108}(24), 240402 (2012).}


\bibitem{Goldberg2020}
\tcr{A.Z. Goldberg, A.B. Klimov, M. Grassl, G. Leuchs, and L.L. S\'{a}nchez-Soto, ``Extremal quantum states,'' AVS Quantum Sci. \textbf{2}(4), 044701 (2020).}


\bibitem{Giraud2008}
\tcr{O. Giraud, P. Braun, and D. Braun, ``Classicality of spin states,'' Phys. Rev. A \textbf{78}(4), 042112 (2008).}

\bibitem{Bouchard2017}
\tcr{F. Bouchard, P. de la Hoz, G. Björk, R. W. Boyd, M. Grassl, Z. Hradil, E. Karimi, A. B. Klimov, G. Leuchs, J. Řeháček, and L. L. Sánchez-Soto,  ``Quantum metrology at the limit with extremal Majorana constellations,''  Optica \textbf{4}(11), 1429--1432 (2017).}

\end{thebibliography}
\end{document}